\documentclass[aps, pra, twocolumn, superscriptaddress]{revtex4}
\usepackage{graphicx}
\usepackage{amsmath}
\usepackage{amssymb}

\usepackage[colorlinks, linkcolor=blue, urlcolor=magenta, citecolor=blue]{hyperref}

\begin{document}

\title{Nonlinear spin dynamics induced by feedback under continuous Larmor frequency distributions}

\author{Tishuo Wang}
\affiliation{Guangdong Provincial Key Laboratory of Quantum Metrology and Sensing, and School of Physics and Astronomy, Sun Yat-Sen University (Zhuhai Campus), Zhuhai 519082, China}
\affiliation{State Key Laboratory of Optoelectronic Materials and Technologies, Sun Yat-Sen University (Guangzhou Campus), Guangzhou 510275, China}

\author{Zhihuang Luo}
\email[]{luozhih5@mail.sysu.edu.cn}
\affiliation{Guangdong Provincial Key Laboratory of Quantum Metrology and Sensing, and School of Physics and Astronomy, Sun Yat-Sen University (Zhuhai Campus), Zhuhai 519082, China}
\affiliation{State Key Laboratory of Optoelectronic Materials and Technologies, Sun Yat-Sen University (Guangzhou Campus), Guangzhou 510275, China}

\author{Shizhong Zhang}
\affiliation{Department of Physics and HKU-UCAS Joint Institute for Theoretical and Computational Physics at Hong Kong, The University of Hong Kong, Hong Kong, China}

\author{Zhenhua Yu}
\email[]{huazhenyu2000@gmail.com}
\affiliation{Guangdong Provincial Key Laboratory of Quantum Metrology and Sensing, and School of Physics and Astronomy, Sun Yat-Sen University (Zhuhai Campus), Zhuhai 519082, China}
\affiliation{State Key Laboratory of Optoelectronic Materials and Technologies, Sun Yat-Sen University (Guangzhou Campus), Guangzhou 510275, China}

\begin{abstract} 
Nonlinear spin dynamics are essential in exploring nonequilibrium quantum phenomena and have broad applications in precision measurement.
Among these systems, the combination of a bias magnetic field and feedback mechanisms can induce self-sustained oscillations at the base Larmor frequency due to nonlinearity. These features have driven the development of single-species and multiple-species spin masers. The latter, with multiple discrete Larmor frequencies, provides significant advantages for precision measurement by mitigating uncertainties in precession frequencies due to long-term drifts in experimental conditions. 
The self-sustained oscillations of single-species and multiple-species spin masers correspond to 
 limit cycles and 
 quasi-periodic orbits of the stable nonlinear dynamics of the systems respectively; the correspondence is elucidated in a recent study on a related spin system featuring two discrete intrinsic Larmor frequencies under dual bias magnetic fields~\cite{wang2023feedback}.
Here, we extend the study to the case that the intrinsic Larmor frequencies of individual spins of the system, given rise to by an inhomogeneous bias magnetic field, form a continuum. We show that generically the stable dynamics of the system includes limit cycles, quasi-periodic orbits, and chaos.
We establish the relation between the synchronization frequency of limit cycles and the field inhomogeneity and derive an equation determining the stability of limit cycles. 
Furthermore, detailed characteristics of different dynamical phases, especially the robustness of limit cycles and  quasi-periodic orbits against experimental fluctuations, are discussed. Our findings not only encompass the case of discrete Larmor frequencies, but also provide crucial insights for precision measurement and the exploration of continuous time crystals and quasi-crystals.

\end{abstract}

\maketitle

\section{Introduction}
Nonlinearity introduces complexity and mystery into the world due to the disproportionate relationship between input and output~\cite{drazin1992nonlinear}. Typical nonlinear dynamical behaviors encompass chaos~\cite{Strogatz2018Nonlinear,alligood2000chaos}, turbulence~\cite{davidson2015turbulence,nazarenko2011wave}, soliton~\cite{Burger1999DarkSolitons,Frantzeskakis_2010}, pattern formation~\cite{zhang2020pattern,Staliunas2002,Kwon2021}, etc.
The nonlinear dynamics is ubiquitous, manifesting across diverse areas, from engineering to science~\cite{Strogatz2018Nonlinear}, few-body to many-body physics~\cite{Musielak_2014,faddeev2013quantum,thouless2013quantum}, and classical to quantum regimes~\cite{falkovich2011fluid,pethick2008bose}.
Spin, as an intrinsic quantum property, plays a crucial role in the foundations of physics and has significant applications, such as quantum computing~\cite{Chuang}, magnetic resonance imaging~\cite{sprawls2000magnetic}, atomic clocks~\cite{RLudlow2015}, magnetometers~\cite{bevington2020magnetic,Gilles03,Jiang2021Floquet,Budker2000}, etc. 
Nonlinearity in spin systems can arise from internal interactions, environmental couplings, or feedback mechanisms, rendering exotic phenomena including time crystals~\cite{greilich2024robust,Phatthamon22,wang2024observationcontinuoustimecrystal,tang2024revealing}, quantum synchronization~\cite{Laskar2020,Roulet2018,Krithika2022}, spin turbulence~\cite{Fujimoto2012Counterflow,Fujimoto2012Spin,Fujimoto2013,Fujimoto2014,Hong2023,Kim2024}, non-thermal fixed points~\cite{Fujimoto2019,lannig2023observation} etc. 

In most cases, spin systems operate in a static bias magnetic field, for example, the representative platforms including alkali-metal vapors \cite{Chalupczak2015Alkali,tang2024revealing}, Rydberg atoms \cite{wu2024dissipative,PhysRevLett.131.143002,ding2023ergodicity}, and nitrogen-vacancy centers \cite{DOHERTY20131}, among others.
Naturally, spins relax to the equilibrium with a constant longitudinal spin polarization. To balance this relaxation, an artificial feedback schemes is introduced, rendering the system nonlinear and producing a limit cycle characterized by stable, persistent precession around the $z$-axis~\cite{Jiang2021Floquet,Yoshimi2002Nuclear,Inoue2016Frequency,PhysRevLett.72.2363,Goldenberg1960Atomic,Oxborrow2012Room,Breeze2018Continuous}, and thus providing a precision frequency reference for precision measurement and probing new physics beyond the standard model~\cite{Rosenberry2001Atomic, Inoue2016Frequency, Bear2000Limit, Safronova2018Search, Terrano2022Comgnetometer,Jiang2021Floquet}. These devices are commonly referred to as spin masers, which can be implemented in various states of matter, from gases \cite{richards19883he,gilles2003he3,Chupp1994Spin} and liquids \cite{Suefke2017Para,marion2008observation,chen2011spontaneous,pravdivtsev2020continuous} to solids \cite{PhysRevLett.38.602,abergel2002self,weber2019dnp,hope2020magic}.
Furthermore, the utilization of two or more species spin masers can effectively mitigate the effects of long-term drifts in the magnetic field and temperature, etc~\cite{Chupp1994Spin, Stoner1996Demonstration, Bear1998Improved, Sato2018Development, Bevington2020Dual}. Since each species experiences the same drifts, their relative oscillation frequencies remain unchanged.
Notably, in most previous works on multiple species masers, each species is treated independently. This is because the gyromagnetic ratios of each species, and therefore their Larmor frequencies, typically differ rather significantly. 
In fact, this is only a special limit in the quasi-periodic orbit phase, as highlighted in a recent study on a single-species spin system in dual bias magnetic fields~\cite{wang2023feedback}. 
The dual bias magnetic fields can be regarded as the simplest case of an inhomogeneous field. However, the spin dynamical behaviors in a more general inhomogeneous bias magnetic field awaits further investigation.

In this work, we systematically study how the inhomogeneity of bias magnetic field affects the performance of the spin systems in a broader parameter space and thus generalize the dual case~\cite{wang2023feedback} to a continuum one. 
Our findings reveal that the phases of limit cycles, quasi-periodic orbits, and chaos are generic.
We derive the equations for the limit cycles determining the relation between $\omega_s$ the synchronization frequency and the field inhomogeneity, and obtain a formula to determine the stability region of the limit cycles. These analytic results are also applicable to the discrete cases and encompass the dual bias magnetic fields case.
As illustrative examples, we analyze two continuum distributions (uniform and root) and determine the stable regions of each dynamical phase through both analytical and numerical methods. We analyze in detail the characteristics of various dynamic phases, and demonstrate the robustness of limit cycles and quasi-periodic orbits against possible experimental noises. Our findings offer valuable insights for the realization of continuous time crystals and quasi-crystals, as well as implications for precision measurement.


\section{Formalism and linear analysis}
We consider an ensemble of $N$ spins placed in an inhomogeneous static bias magnetic field 
applied along the $z$-axis, resulting in that the Larmor frequency for the $j$th spin is $\omega_j$; in the continuum limit, the distribution $\rho(\omega)\equiv \frac1N\sum_{j=1}^N\delta(\omega-\omega_j)$ for the Larmor frequencies is a smooth function.
The expectation value of a spin with a specific Larmor frequency $\omega$ is represented by $\mathbf P(\omega,t)=\{ P_x(\omega,t),P_y(\omega,t),P_z(\omega,t)\}$. We take $\hbar=1$ throughout such that $\mathbf P(\omega,t)$ is dimensionless. Moreover, a feedback circuit is introduced to use the average spin polarization $\overline{\mathbf P}(t)=\int_{-\infty}^{\infty}d\omega\rho(\omega)\mathbf P(\omega,t)$ measured to generate a transverse magnetic field $\mathbf B^{\text{fb}}(t)=\{\frac{\alpha}{\gamma}\overline P_{y}(t),-\frac{\alpha}{\gamma}\overline P_{x}(t),0\}$ with $\alpha$ the amplification factor and $\gamma$ the gyromagnetic ratio of the spins \cite{Yoshimi2002Nuclear,Sato2018Development,Inoue2016Frequency}.  The dynamics of $\mathbf P(\omega,t)$ can be described by the nonlinear Bloch equations
\begin{align}
	\frac{d P_{x}(\omega,t)}{dt}=&\omega P_{y}(\omega,t)+\alpha \overline{P}_x(t) P_{z}(\omega,t)-\frac{P_{x}(\omega,t)}{T_2},\label{blochxCon}\\
	\frac{d P_{y}(\omega,t)}{dt}=&-\omega P_{x}(\omega,t)+\alpha \overline{P}_y(t) P_{z}(\omega,t)-\frac{P_{y}(\omega,t)}{T_2},\label{blochyCon}\\
	\frac{d P_{z}(\omega,t)}{dt}=&-\alpha \left [ \overline{P}_x(t)P_{x}(\omega,t)+\overline{P}_y(t) P_{y}(\omega,t)\right ]\nonumber\\
	&-\frac{P_{z}(\omega,t)-P_0}{T_1}\label{blochzCon},
\end{align}
where  $T_2$ represents the spin transverse relaxation time, and $T_1$ and $P_0$ are respectively the effective longitudinal relaxation time and the equilibrium polarization, with the effect of pumping taken into account~\cite{tang2024revealing,Happer1984Polarization,Happer1972Optical,Walker1997Spin}.

%
The full stable dynamics of Eqs.~(\ref{blochxCon}) to (\ref{blochzCon}) is given below. Since the present continuum limit shall retrieve the single-species case when $\rho(\omega)\sim\delta(\omega-\omega_0)$ with $\omega_0$ the uniform Larmor frequency for all spins \cite{Yoshimi2002Nuclear,Inoue2016Frequency}, limit cycles are expected to be a stable solution to the system in certain parameter regimes; actually we can derive useful analytic results for limit cycles. 

\begin{figure}[t]
	\centering
	\includegraphics[width=0.5\textwidth]{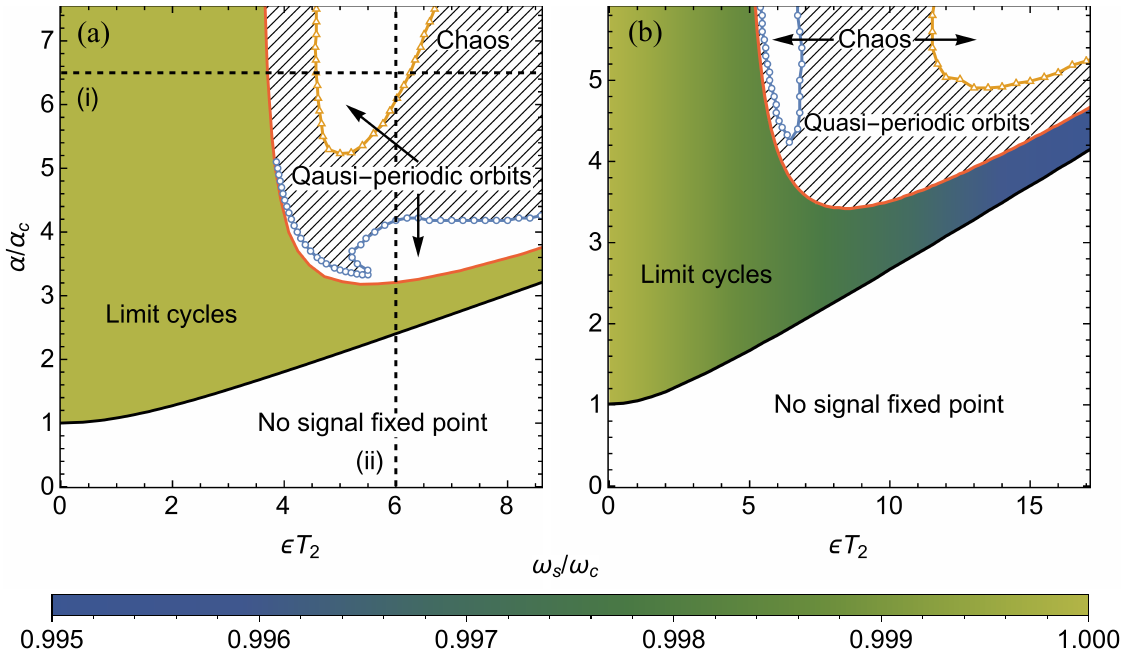}
	\caption{Stability diagrams for (a) the uniform distribution $\rho_U(\omega)$ and (b) the root distribution $\rho_R(\omega)$, showing phases of no signal fixed point, limit cycles, quasi-periodic orbits and chaos. Here $\alpha_c\equiv1/T_2P_0$ is the critical amplification factor for the case of a single Larmor frequency (corresponding to $\epsilon=0$)~\cite{Yoshimi2002Nuclear, Inoue2016Frequency}. 
	The solid curves agree with Eqs.~(\ref{m}) and~(\ref{beta}). The symbols of empty circles and triangles mark the boundaries calculated numerically.  
		The shading of the limit cycle phase represents the magnitude of the synchronization frequency $\omega_s$. Along the dashed lines (i) and (ii) in (a), the largest Lyapunov exponents are calculated and shown in Fig.~\ref{fig: LyExponent}.}
	\label{fig: PhasediagramContinuum}
\end{figure}

To work out properties of the limit cycle solution, we transform the system to a rotating frame of frequency $\omega_s$ (to be determined self-consistently below) as $\tilde{P}_T(\omega,t)=e^{i\omega_s t} P_T(\omega,t)$ and $\tilde{\overline P}_T=e^{i\omega_s t} \overline P_T$, where $P_T(\omega,t)\equiv P_x(\omega,t)+iP_y(\omega,t)$. By requiring $d\tilde{P}_T(\omega,t)/dt=0$ and $d{P}_z(\omega,t)/dt=0$, and making use of the self-consistent condition $\tilde{\overline P}_T=\int_{-\infty}^{\infty}d\omega\rho(\omega)\tilde P_T(\omega,t)$, we find 

\begin{align}
	&0=\int_{-\infty}^{\infty}(\omega-\omega_s)\rho(\omega)\mathcal{L}(\omega) d\omega,
	\label{os}\\
	&\frac{T_1}{\alpha P_0T_2}=\int_{-\infty}^{\infty}\rho(\omega)\mathcal{L}(\omega) d\omega,\label{tbpt}
\end{align}
where $\mathcal{L}(\omega) \equiv \{[1+(\omega-\omega_s)^2T_2^2]/T_1+\alpha^2 T_2|\tilde{\overline P}_T|^2\}^{-1}$.
Together, these two equations determine the synchronization frequency $\omega_s$ and the amplitude $\tilde{\overline P}_T$ (apart from an arbitrary phase). 
Consequently the limit cycle solution can be written as 
\begin{align}
& P_T(\omega,t)=\frac{\alpha P_z(\omega)\tilde{\overline P}_T }{1/T_2+i(\omega-\omega_s)}e^{-i\omega_s t},\label{tildePT}\\
&P_z(\omega)=P_0[1+(\omega-\omega_s)^2T_2^2]\mathcal{L}(\omega)/T_1. \label{Pz}
\end{align}
According to Eq.~(\ref{os}), if $\rho(\omega,t)$ is a symmetric function with respect to a frequency $\tilde\omega$, the synchronized frequency $\omega_s$ shall equal $\tilde\omega$; otherwise, $\omega_s$ shall change when $\alpha$ is tuned (see, for example, Fig.~\ref{fig: PhasediagramContinuum}). 

The stability of the above limit cycle solution can be analyzed by assuming
the perturbations $\delta\tilde P_T(\omega,t)=e^{\beta t}F(\omega)$ and $\delta P_z(\omega,t)=e^{\beta t}G(\omega)$ around the solution, Eqs.~(\ref{tildePT}) to (\ref{Pz}). Here $F(\omega)$ is a complex function and $G(\omega)$ is a real one and $\beta$ represents the growth rate. 
In the linear regime, these perturbations satisfies the linearized Bloch equations 
\begin{align}
	\mathcal M
	\begin{bmatrix} \delta P_z(\omega,t) \\ \delta\tilde P_T(\omega,t) \\ \delta\tilde P_T^*(\omega,t)\end{bmatrix}
	=\begin{bmatrix} -\alpha [\delta\tilde{\overline P}_T^*(\omega,t) \tilde P_T(\omega)+\delta\tilde{\overline P}_T(\omega,t)\tilde P_T^*(\omega)]/2\\
		\alpha P_z(\omega) \delta\tilde{\overline P}_T(\omega,t)\\
		\alpha P_z(\omega) \delta\tilde{\overline P}_T^*(\omega,t)\end{bmatrix} \label{bm},
\end{align}
where
\begin{align}
	&\mathcal M=\nonumber\\
	&\begin{bmatrix} \beta+\frac1{T_1} & \alpha\tilde{\overline P}_T^*/2 &  \alpha\tilde{\overline P}_T/2\\
		-\alpha\tilde{\overline P}_T & \beta+\frac1{T_2}+i(\omega-\omega_s) & 0\\
		-\alpha\tilde{\overline P}_T^* & 0 &  \beta+\frac1{T_2}-i(\omega-\omega_s)\end{bmatrix},\label{m}
\end{align}
Via Eq.~(\ref{bm}), we express $\delta\tilde P_T(\omega,t)$ and $\delta\tilde P_T^*(\omega,t)$ in terms of $\delta\tilde{\overline P}_T(\omega,t)$ and $\delta\tilde{\overline P}_T^*(\omega,t)$. By utilizing $\delta\tilde{\overline P}_T(\omega,t)=\int_{-\infty}^\infty d\omega\rho(\omega)\delta\tilde P_T(\omega,t)$, we derive the characteristic equation
\begin{widetext}
	\begin{align}
		&\left\{1-\alpha \int_{-\infty}^{\infty}d\omega\rho(\omega) [\mathcal M^{-1}_{22}P_z(\omega)-\mathcal M^{-1}_{21}\tilde P_T^*(\omega)/2]\right\}
		\left\{1-\alpha \int_{-\infty}^{\infty}d\omega\rho(\omega) [\mathcal M^{-1}_{33}P_z(\omega)-\mathcal M^{-1}_{31}\tilde P_T(\omega)/2]\right\}\nonumber\\
		&=\alpha^2 \left\{\int_{-\infty}^{\infty}d\omega\rho(\omega) [\mathcal M^{-1}_{23}P_z(\omega)-\mathcal M^{-1}_{21}\tilde P_T(\omega)/2]\right\}
		\left\{\int_{-\infty}^{\infty}d\omega\rho(\omega) [\mathcal M^{-1}_{32}P_z(\omega)-\mathcal M^{-1}_{31}\tilde P_T^*(\omega)/2]\right\}.\label{beta}
	\end{align}
\end{widetext}
which determines the growth rate $\beta$, with $\mathcal M^{-1}$ being the inverse matrix of $\mathcal M$.
The limit cycles become unstable when any solution for 
$\beta$ has a positive real part. Notably, setting $\tilde{\overline P}_T$ to zero simplifies Eqs.~(\ref{m}) and~(\ref{beta}) to yield the stability criterion for the no signal fixed point. To study stable dynamics of Eqs.~(\ref{blochxCon}) to (\ref{blochzCon}) other than limit cycles and no signal fixed point, we shall turn to numerical methods.

\begin{figure*}[htbp]
	\centering
	\includegraphics[width = 0.9\linewidth]{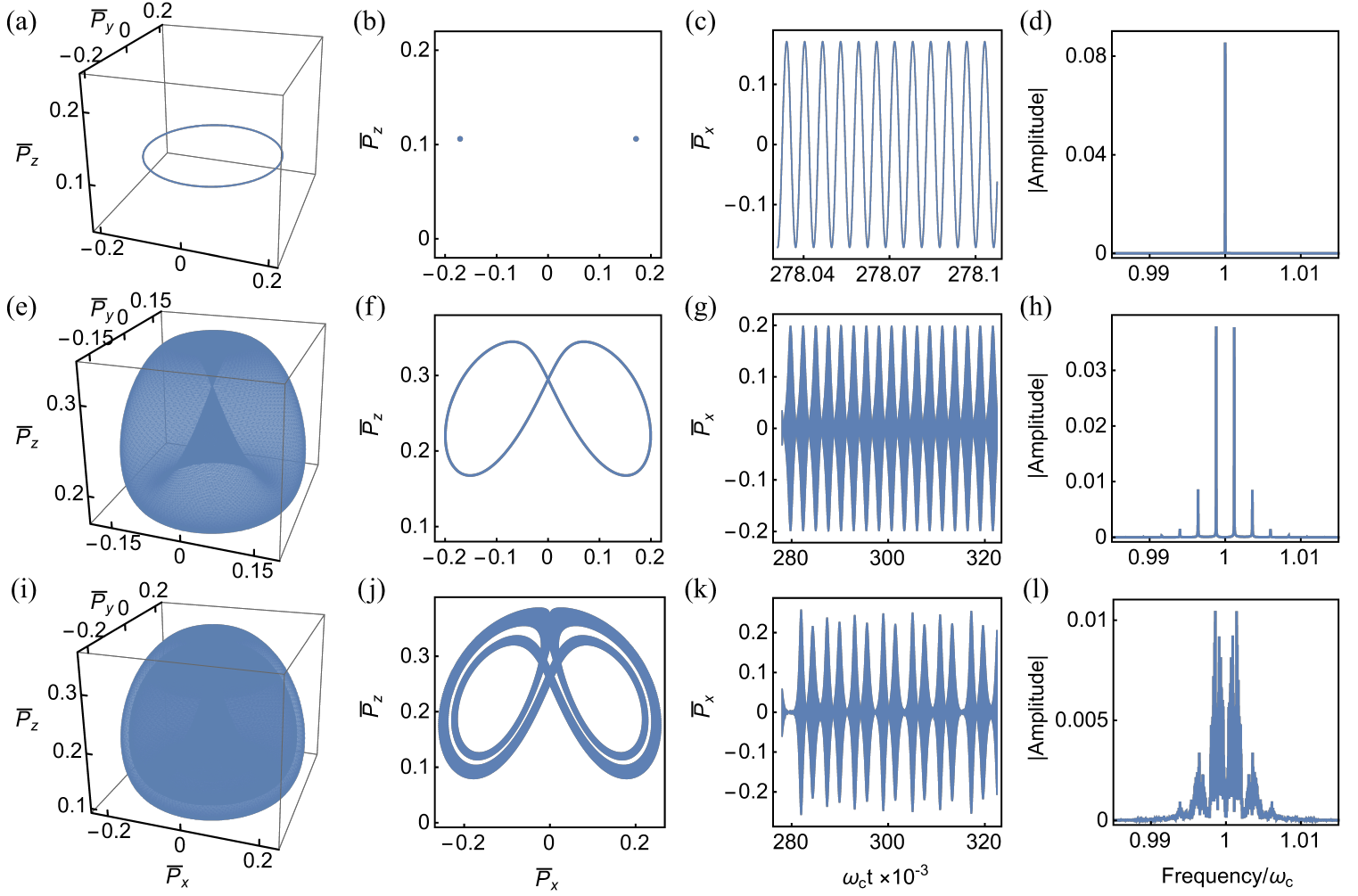}
	\caption{
	Representative dynamical behaviors of limit cycles (the top row), quasi-period orbits (the middel row) and chaos (the bottom row) for the uniform distribution $\rho_U(\omega)$. The first column displays the trajectories of the average spin polarization $\overline{\mathbf P}(t)$, while the second column shows  the trajectory cross-sections at $\overline{P}_{y}=0$. The third column depicts the time series of $\overline{P}_x(t)$, and the fourth column presents the absolute value of the single-side Fourier transform amplitude of $\overline{P}_x(t)$. The parameters for the plots are $\alpha/\alpha_c=4$, $\epsilon T_2=1$ for (a)-(d), $\alpha/\alpha_c=4$, $\epsilon T_2=6$ for (e)-(h), and $\alpha/\alpha_c=4$, $\epsilon T_2=5$ for (i)-(l). }
	\label{fig: phaseportrait}
\end{figure*}

\section{Numerical results}

To calculate stable dynamics of Eqs.~(\ref{blochxCon}) to (\ref{blochzCon}) numerically, we assume experimentally realistic parameters $P_0=0.392097$, $T_1=13.0699 \rm s$, $T_2=13.65 \rm s$ \cite{Jiang2021Floquet}, and consider two following exemplary distributions for $\rho(\omega)$ nonzero in the vicinity of the frequency $\omega_c/2\pi=8.85 \rm Hz$. More details on the numerical treatment are given in Appendix A.


In the presence of a bias magnetic field with a linear gradient, if the spins are distributed uniformly in the real space, the distribution $\rho(\omega)$ for the Larmor frequencies is given 
 by
\begin{align}
	\rho_U(\omega)=[\theta(\omega-\omega_c+\epsilon/2)-\theta(\omega-\omega_c-\epsilon/2)]/\epsilon,
\end{align}
where $\theta$ is the Heaviside function and $\epsilon$ quantifies the width of the distribution. We call $\rho_U(\omega)$ the uniform distribution.

The other example is to consider uniform spatial distribution of spins inside a small cylindrical cell with radius $r$ and height $h$, aligned with its symmetric axis along the $\hat z$-axis and placed at the center with respect to a pair of identical Helmholtz coils. 
The magnetic field $B_z$ produced by the coils around the center (the origin) has the approximate expression, $B_z\approx[\omega_c+a z^2+ b (x^2+y^2)]/\gamma$~\cite{purcell2013electricity}, where $\omega_c$, $a$ and $b$ are parameters determined by the properties of the coils. 
For simplicity, we assume $a h^2 = -2 b r^2$ ($\geq 0$), and obtain the distribution
\begin{align}
	\rho_R(\omega)=\begin{cases}
  \frac{3}{ 2\epsilon}\sqrt{\frac{3(\omega-\omega_c)+2\epsilon}{ \epsilon }} ,\quad &1-\frac{2\epsilon}{3\omega_c}\le \frac{\omega}{\omega_c}< 1-\frac{\epsilon}{3\omega_c} \\
   \frac{3}{2 \epsilon},\quad  & 1-\frac{\epsilon}{3\omega_c}\le \frac{\omega}{\omega_c}< 1 \\
 \frac{3}{2\epsilon}(1-\sqrt{\frac{3(\omega-\omega_c)}{ \epsilon}}) ,\quad  & 1\le \frac{\omega}{\omega_c}\leq 1+\frac{\epsilon}{3\omega_c}
  \end{cases}
  \end{align}
with the distribution width $\epsilon=3a h^2/4$. We call $\rho_R(\omega)$ the root distribution.

\subsection{Stability diagram}

The corresponding stability diagram for $\rho_U(\omega)$ and $\rho_R(\omega)$ are presented in Fig.~\ref{fig: PhasediagramContinuum}(a) and (b) respectively. 
The limit cycles, quasi-periodic orbits and chaos are found to be the general stable dynamics of the system, exhibiting stability in different regions.
The boundaries represented by the solid lines agree with Eqs.~(\ref{m}) and~(\ref{beta}), while the other boundaries marked by symbols are determined numerically. For the uniform distribution $\rho_U(\omega)$, the synchronization frequency $\omega_s$, represented by the shading color values, is equal to $\omega_c$ throughout the entire limit cycle region. In contrast,  for the root distribution $\rho_R(\omega)$, $\omega_s$ varies across the stable region of the limit cycles. As $\epsilon$ and $\alpha$ vary, the system may enter the stable regions of the quasi-periodic orbits and chaos. 


In Fig.~\ref{fig: phaseportrait}, we show the stable dynamical behaviors across different phases for the uniform distribution $\rho_U(\omega)$ by taking $\alpha/\alpha_c=4$ and varying $\epsilon$. 
We present the phase portrait in the phase space of $\overline{\mathbf{P}}(t)$ (the first column), which intersects with $\overline{P}_y=0$ giving rise to a Poincaré section (the second column). 
Alongside the signal (third column) and the corresponding absolute value of Fourier transform amplitude (fourth column), we observe that in the limit cycle phase (the top row), the signal $\overline{P}_x(t)$ oscillates sinusoidally  at the frequency $\omega_s$,  consistent with the predictions from Eqs.~(\ref{os}) and~(\ref{tbpt}). 
As $\epsilon$ increases, the system transits into the quasi-periodic orbits (the middle row) or the chaos (the bottom row),
and the trajectories of  $\overline{\mathbf{P}}(t)$ appear quite dense, as illustrated in Fig.~\ref{fig: phaseportrait}(e) and (i). However, the distinction between the two phases is immediately obvious in the corresponding Poincaré sections:  the intersecting points for the quasi-periodic orbits form a smooth closed curve, while those for the chaos cluster into dense bands. Additionally, the Fourier spectrum of $\overline P_x (t)$ peaks at regular frequencies for the quasi-periodic orbits, whereas in the chaotic regime, the peaks exhibit irregularity.
\begin{figure}[htbp]
	\centering
	\includegraphics[width=0.48\textwidth]{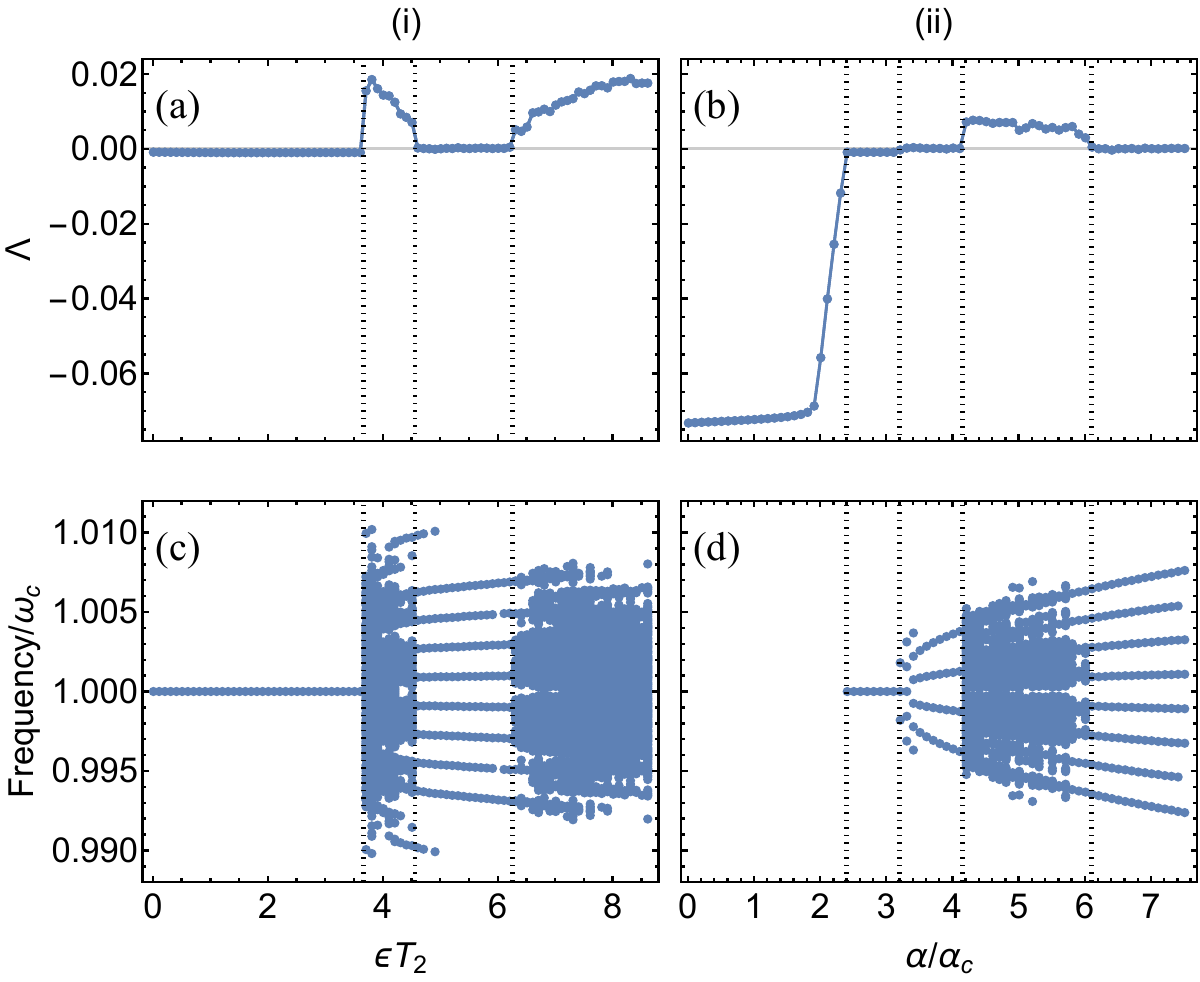}
	\caption{Top row: the largest Lyapunov exponents for the parameters tuned as denoted by (i) and (ii) in Fig.~\ref{fig: PhasediagramContinuum}(a), calculated using the algorithm outlined in Appendix B.  For case (i), $\alpha/\alpha_c = 6.5$, and for case (ii), $\epsilon T_2 = 6$. Bottom row: the corresponding Fourier spectrum analysis showing the frequencies at which the absolute value of the Fourier transform amplitude of $\overline P_x(t)$ peaks.  }
	\label{fig: LyExponent}
\end{figure} 

Trajectories in the chaos phase are known to be sensitive to the initial states. Such a sensitivity may have a prospect of application in precision measurement. Another quantitative method to distinguish chaos from other types of attractors is through the largest Lyapunov exponent, $\Lambda$, which represents the maximum exponential rate at which two nearby initial states in the vicinity of an attractor evolve to diverge~\cite{Strogatz2018Nonlinear}. A positive $\Lambda$ indicates chaos, a negative value corresponds to fixed points, and $\Lambda = 0$ for limit cycles and quasi-periodic orbits respectively~\cite{HAKEN198371, parker2012practical,sandri1996numerical}. Figure~\ref{fig: LyExponent}(a) and (b) show the largest the largest Lyapunov exponent for the uniform distribution $\rho_U(\omega)$. These results are consistent with both the corresponding Fourier spectrum analysis in  Fig.~\ref{fig: LyExponent} (c)-(d) and the stability diagram illustrated in Fig.~\ref{fig: PhasediagramContinuum}(a).


\subsection{Robustness of limit cycles and quasi-periodic orbits}
\begin{figure}
	\centering
	\includegraphics[width=0.48\textwidth]{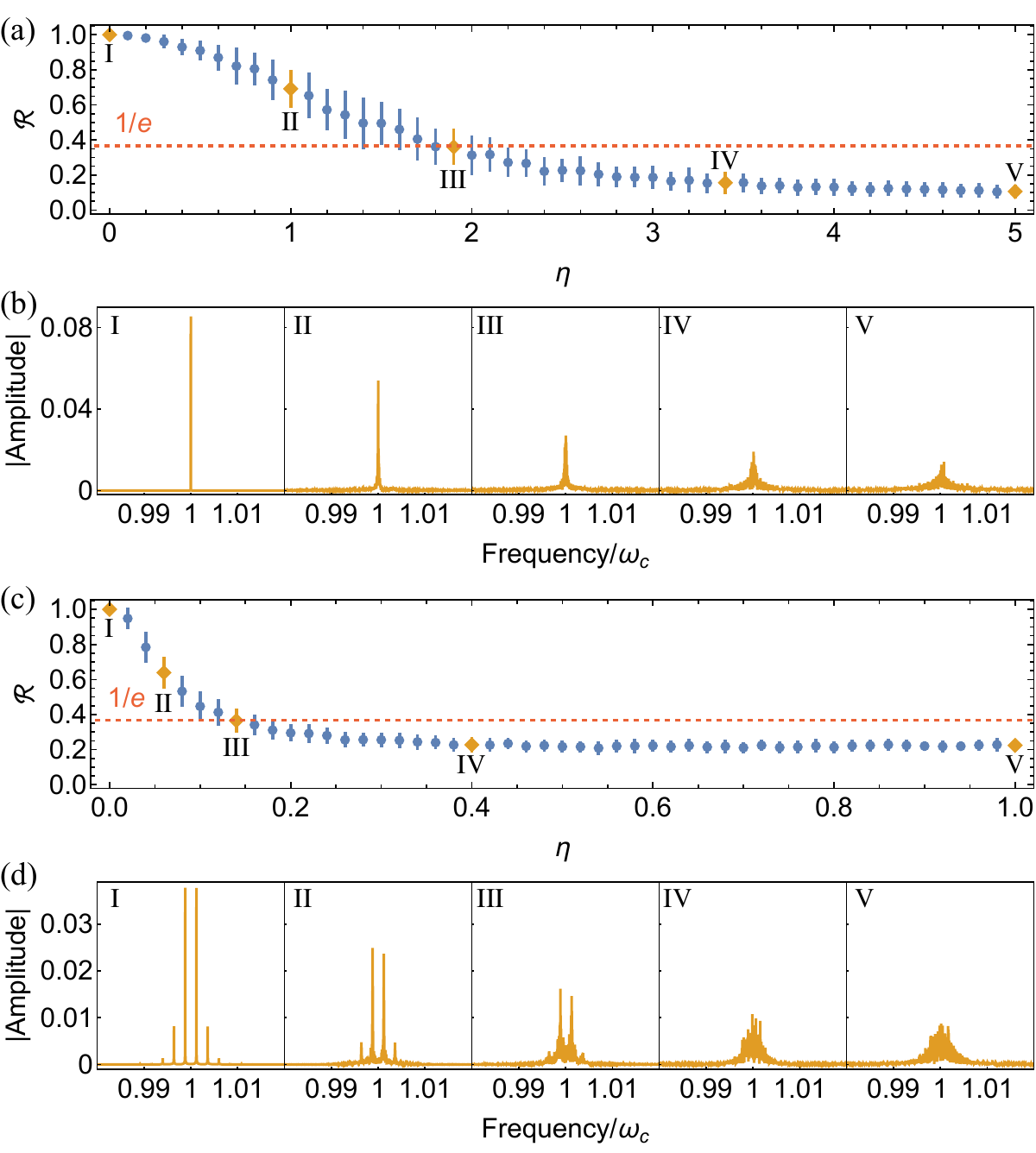}
	\caption{Robustness of the limit cycles [(a)-(b)], quasi-periodic orbits [(c)-(d)] for the uniform distribution against the noises $l_\eta(t)$ and $g_\eta(t)$. (a) and (c) plot the function~(\ref{eq: robust}), where the error bars correspond to the standard deviations from $50$ individual simulations. (b) and (d) show the single-shot Fourier transform spectra of $\overline P_x(t)$ for different $\eta$, marked by the diamond symbols in (a) and (c). Parameters taken here are $\alpha/\alpha_c=4$, $\epsilon T_2=1$ for (a)-(b) and $\alpha/\alpha_c=4$, $\epsilon T_2=6$ for (c)-(d).}
	\label{fig: probustB}
\end{figure}

\begin{figure}
	\centering
	\includegraphics[width=0.48\textwidth]{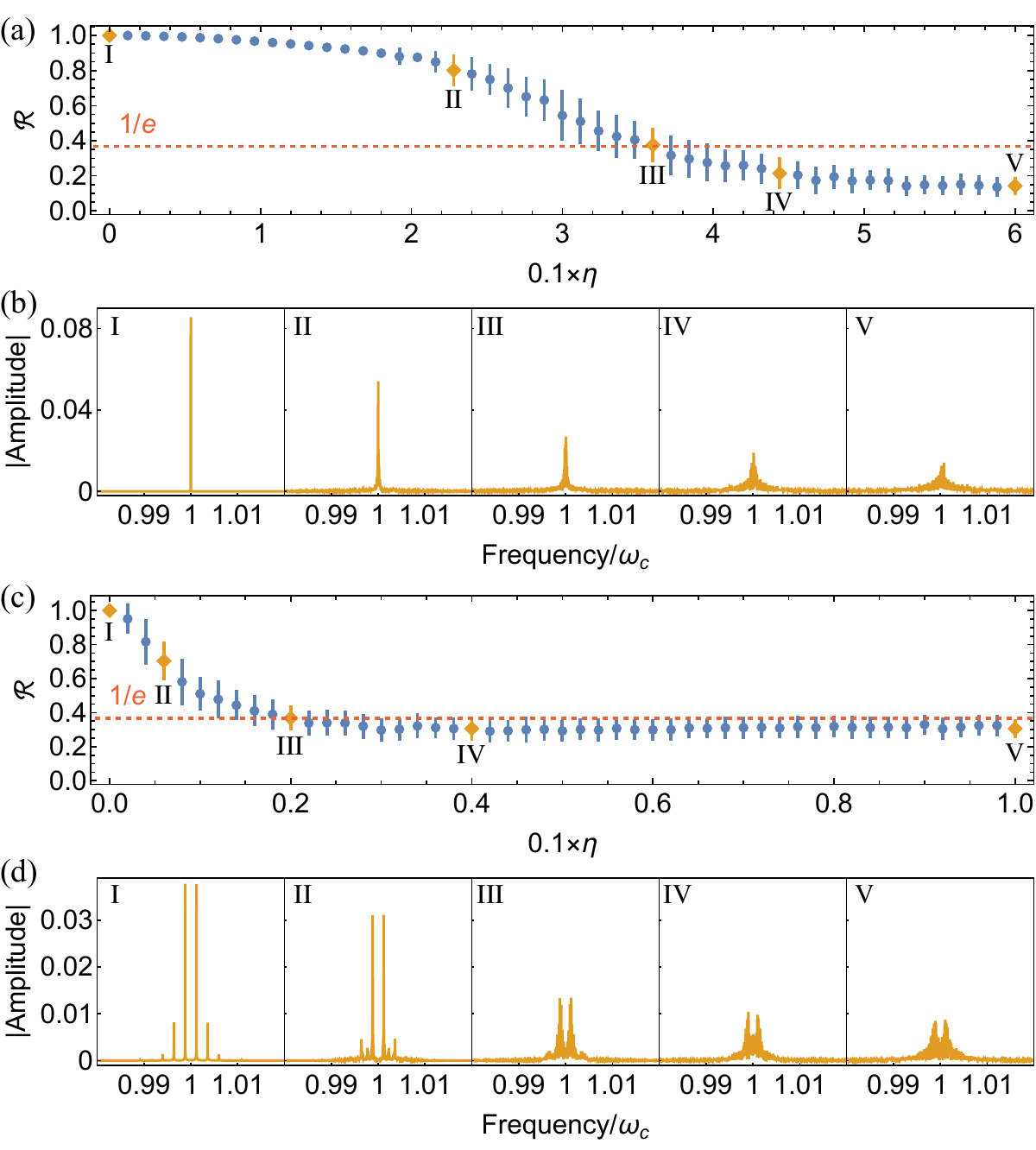}
	\caption{Robustness of the limit cycles [(a)-(b)], quasi-periodic orbits [(c)-(d)] for the uniform distribution against the noise $s_\eta(t)$. The quantities plotted and parameters used are the same as those in Fig.~\ref{fig: probustB}.}
	\label{fig: probustAlpha}
\end{figure}
The prominent peaks in the frequency spectrum during the limit cycle and quasi-periodic phases make them highly useful for precision measurement \cite{Jiang2021Floquet,Sato2018Development,Inoue2016Frequency}. However, it is crucial to demonstrate the robustness of these phases against noises inevitable in realistic implementations. Noises can stem from various sources, such as feedback electronics, bias fields, etc.

We model any noisy field $l_\eta(t)$ added on top of an original ideal field $l_0(t)$ as a function generating random numbers within the range $[-\eta,\eta]$ at any time $t$, where $\eta$ represents the strength of the noise. Once with a noise introduced into Eqs.~(\ref{blochxCon}) to (\ref{blochzCon}), the stable dynamics of any observable $p_\eta(t)$ will differ from that with $\eta=0$. To quantify this difference, we define the function
\begin{align}
\mathcal R & = \frac{\int  |\tilde p_0(f)\tilde p_\eta(f)| df}{\sqrt{\int |\tilde p_0(f)|^2 df \int |\tilde p_\eta(f')|^2 df'}}.
	\label{eq: robust}
\end{align}
where $\tilde p_\eta(f)$ is the Fourier transform amplitude of $p_\eta(t)$. Here we consider noises originating from two distinct sources separately: first, the feedback electronics, which yields the noisy transverse fields as $\mathbf B_{\mathrm{noise}}^{\text{fb}}(t)=\{[\alpha\overline P_{y}(t)+l_\eta(t)]/\gamma, -\alpha\overline P_{x}(t)+g_\eta(t)]/\gamma,0\}$; second, the electronic apparatus for amplification, 
which leads the amplification factor to be $\alpha_{\mathrm{noise}}(t)=\alpha+s_\eta(t)$, and consequently $\mathbf B_{\mathrm{noise}}^{\text{fb}}(t)=\{[\alpha\overline P_{y}(t)+s_\eta(t)\overline P_{y}(t)]/\gamma, -\alpha\overline P_{x}(t)-s_\eta(t)\overline P_{x}(t)]/\gamma,0\}$. Note that $l_\eta(t)$ and $g_\eta(t)$ are independent, and $\overline P_{x}(t)$ and $\overline P_{y}(t)$ are dimensionless. 

Figure~\ref{fig: probustB} and \ref{fig: probustAlpha} show the robustness of the limit cycles and quasi-periodic orbits under the influence of  $l_\eta(t)$ and $g_\eta(t)$, and $s_\eta(t)$ respectively. 
We define $1/e$ as the threshold value of $\mathcal{R}$~\cite{Phatthamon22}. In both the cases, the limit cycles and quasi-periodic orbits exhibit robustness over a substantial range of $\eta$, as also evidenced by the single-shot spectra displayed in the second and fourth rows of Figs.~\ref{fig: probustB} and \ref{fig: probustAlpha}.
Though, on the face of it, $\mathcal R$ looks to decrease more slowly against $\eta$ in the case of $s_\eta(t)$ compared with that of $l_\eta(t)$ and $g_\eta(t)$. According to the expression of $\mathbf B_{\mathrm{noise}}^{\text{fb}}(t)$ given above, one can largely consider $s_\eta(t)\overline P_{y}(t)$ and $-s_\eta(t)\overline P_{x}(t)$ equivalent to $l_\eta(t)$ and $g_\eta(t)$. 
Since $\sqrt{\langle \overline P^2_{x,y}(t)\rangle}\sim 0.1$ (see the third column of Fig.~\ref{fig: phaseportrait}), the difference in the dependence of $\mathcal R$ on $\eta$ can be mostly understood.
Relatively speaking, due to the presence of multiple peaks in the spectra, the quasi-periodic orbits are more sensitive to the noises than the limit cycles, resulting in a smaller robust region.

\section{Conclusion}
As a generalization of our previous work \cite{wang2023feedback},  we have studied the feedback assisted nonlinear spin system in an inhomogeneous bias magnetic field resulting in a continuum distribution of intrinsic Larmor frequencies. We establish the relation between the synchronization frequency of limit cycles and the field inhomogeneity, i.e., Eqs.~(\ref{os}) and~(\ref{tbpt}), and derive the stability conditions of the limit cycles, Eqs.~(\ref{m}) and~(\ref{beta}). 
    These results can be used to quantify field inhomogeneity in precision measurement.

We calculate the full stability diagram for the uniform distribution and root distribution as demonstrations. We find that limit cycles, quasi-periodic orbits and chaos are the general stable dynamical phases of the system. In addition,
the robustness of the limit cycle and quasi-periodic orbit phases against noises are evaluated, which is important in the applications of engineering novel single-mode and multimode spin masers~\cite{Yoshimi2002Nuclear,Sato2018Development,Jiang2021Floquet,ChackoMultimode2024}, as well as exploring possible time crystals and quasi-crystals~\cite{greilich2024robust,Phatthamon22,wang2024observationcontinuoustimecrystal,tang2024revealing,wu2024dissipative,PhysRevLett.131.143002,yang2024emergent,Autti2018,Giergiel2018,Huang2018Symmetry-breaking,he2024experimental}. 
Moreover, our formalism can retrieve previous cases with discrete multiple Larmor frequencies; for example, considering $\mathcal N$ cells of spins where 
the $j$th cell is subject to a bias field $B_{j,z}$, the distribution simply takes the form $\large\rho(\omega)=\frac1{\mathcal N}\Sigma_j^{\mathcal N} \delta(\omega-\gamma B_{j,z})$~\cite{wang2023feedback}.

 



\section{Acknowledgements}
We thank Xiaodong Li and Long Wang for their discussions on numerics. TW and ZY are supported by the National Natural Science Foundation of China Grant No.12074440, and Guangdong Project (Grant No.~2017GC010613). ZL is supported by the National Natural Science Foundation of China Grant No. 
11805008, Fundamental Research Funds for the Central Universities, Sun-Yat-Sen University (Grant No. 23qnpy63), and Guangdong Provincial Key Laboratory (Grant No. 2019B121203005). SZ is supported by grants from the Research Grants Council of the Hong Kong Special Administrative Region, China (HKU 17304719, HKU C7012-21GF, and a RGC Fellowship award HKU RFS2223-7S03).
\section*{APPENDIX}
\subsection{Numerical treatment of the continuum limit}
To numerically investigate the nonlinear dynamics of the continuum system $\mathbf P(\omega,t)$, we discretize the average spin polarization as
\begin{align}
\overline{\mathbf P}(t)&=\int_{-\infty}^{\infty}d\omega\rho(\omega)\mathbf P(\omega,t) \nonumber\\
&=\sum_{\nu=1}^{M-1}[\mathbf P(\omega_\nu,t)\rho(\omega_\nu)+\mathbf P(\omega_{\nu+1},t)\rho(\omega_{\nu+1})]\delta\omega/2,
\end{align} 
with $\omega_1$ and $\omega_{M}$ the minimum and maximum values of the Larmor frequency continuum. The width of the continuum is $\epsilon\equiv \omega_{M}-\omega_1$, and the interval is $\delta\omega\equiv\epsilon/(M-1)$.
We then turn to integrate the following $3M$ discrete ordinary differential equations (ODEs)
\begin{align}
	\frac{d P_{x}(\omega_\nu,t)}{dt}=&\omega_\nu P_{y}(\omega_\nu,t)+\alpha \overline{P}_x(t) P_{z}(\omega_\nu,t)-\frac{P_{x}(\omega_\nu,t)}{T_2},\label{blochxConj}\\
	\frac{d P_{y}(\omega_\nu,t)}{dt}=&-\omega_\nu P_{x}(\omega_\nu,t)+\alpha \overline{P}_y(t) P_{z}(\omega_\nu,t)-\frac{P_{y}(\omega_\nu,t)}{T_2},\label{blochyConj}\\
	\frac{d P_{z}(\omega_\nu,t)}{dt}=&-\alpha \left [ \overline{P}_x(t)P_{x}(\omega_\nu,t)+\overline{P}_y(t) P_{y}(\omega_\nu,t)\right ]\nonumber\\
	&-\frac{P_{z}(\omega_\nu,t)-P_0}{T_1}\label{blochzConj}.
\end{align}
Throughout this work, we have set $M=81$.

\subsection{Algorithm for calculating the largest Lyapunov exponent}
The largest Lyapunov exponent $\Lambda$ characterizes how the nearby trajectories in the vicinity of an attractor diverge or converge from each other~\cite{Strogatz2018Nonlinear}. To show how to calculate $\Lambda$~\cite{parker2012practical,sandri1996numerical,Benettin1976}, we first rewrite Eqs.~(\ref{blochxConj})-(\ref{blochzConj}) in shorthand as $\dot{\mathbf x}=\mathbf F(\mathbf x)$ with $ \mathbf x = \{\mathbf P(\omega_j,t)\}$ and express the trajectories from an initial state $\mathbf x(t=0)=\mathbf x_0$ as $ \mathbf x (t) = \mathbf f(\mathbf x_0,t)$. Next, considering two nearby initial states
$\mathbf x_0$ and $\mathbf x_0+\delta \mathbf x^{(0)} $ where $\delta \mathbf x^{(0)}$ is a very small separation, after a short time $\tau$, 
the separation evolves into 
\begin{align}
	\delta \mathbf x^{(1)} &= \mathbf f (\mathbf x_0+\delta \mathbf x^{(0)},\tau)-\mathbf f(\mathbf x_0,\tau)\overset{\delta \mathbf x^{(0)}\to0}{=}D_{\mathbf{x}_0}\mathbf{f}\cdot\delta \mathbf x^{(0)}\nonumber\\
	&\equiv\mathbf{\Phi}(\mathbf{x}_0,\tau)\cdot\delta \mathbf x^{(0)},
\end{align}
and the normalized separation vector $\mathbf v^{(1)}$ becomes
\begin{align}
	\mathbf v^{(1)}&\equiv\frac{\delta \mathbf x^{(1)}}{|\delta \mathbf x^{(0)}|}=\mathbf{\Phi}(\mathbf{x}_0,\tau)\cdot\mathbf{u}^{(0)},
\end{align}
with $\mathbf{u}^{(0)}=\delta \mathbf x^{(0)}/|\delta \mathbf x^{(0)}| $. Similarly, at the $i$th time step, i.e., $t=i\tau$, we have
\begin{align}
\mathbf v^{(i+1)}&=\mathbf{\Phi}(\mathbf{x}_i,t)\cdot\mathbf{u}^{(i)},
\end{align}
where $\mathbf x_i=\mathbf f(\mathbf x_{i-1},\tau)$ and $\mathbf u^{(i)}=\mathbf v^{(i)}/|\mathbf v^{(i)}|$. Then after performing this iteration a sufficient number of times $K$, the largest Lyapunov exponent is given by 
\begin{align}
	\Lambda\approx\frac{1}{K\tau}\sum_{i=1}^K\ln|\mathbf v^{(i)}|.
\end{align}

Here $\mathbf{\Phi}(\mathbf{x}_0,t)$ obeys the variational equation
\begin{align}
	\dot{\mathbf{\Phi}}(\mathbf{x}_0,t)
	&=D_\mathbf{x}\mathbf{F}\cdot\mathbf{\Phi}(\mathbf{x}_0,t), \\
	\mathbf{\Phi}(\mathbf{x}_0,0)& =\mathbf{I}, 
\end{align}
where $D_\mathbf{x}\mathbf{F}$ is the Jacobi matrix at $ \mathbf x (t) = \mathbf f(\mathbf x_0,t)$. Therefore $\mathbf{\Phi}(\mathbf{x}_0,t)$ can be numerically solved by combining with Eqs.~(\ref{blochxConj})-(\ref{blochzConj})
\begin{align}
	\begin{bmatrix}\dot{\mathbf x}\\\dot{\mathbf \Phi}\end{bmatrix}=\begin{bmatrix}\mathbf F(\mathbf x)\\D_{\mathbf x}\mathbf F(\mathbf x)\cdot\mathbf\Phi\end{bmatrix},\quad\begin{bmatrix}\mathbf x(t_0)\\\mathbf \Phi(t_0)\end{bmatrix}=\begin{bmatrix}\mathbf x_0\\\mathbf{I} \end{bmatrix}.
	\label{eq: x and Phi 1}
\end{align}
This formulation allows for the simultaneous numerical integration of both the system's equations of motion and the evolution of $\mathbf \Phi $, enabling the calculation of the largest Lyapunov exponent.

\bibliographystyle{apsrev4-1}
\bibliography{Ref_maser}

\end{document}